# Comparison of Deep Learning and Machine Learning Models and Frameworks for Skin Lesion Classification

Soham Bhosale

July 26, 2022

**Abstract:**

The incidence rate for skin cancer has been steadily increasing throughout the world, leading to it being a serious issue[1]. Diagnosis at an early stage has the potential to drastically reduce the harm caused by the disease, however, the traditional biopsy is a labor-intensive and invasive procedure[7]. In addition, numerous rural communities do not have easy access to hospitals and do not prefer visiting one for what they feel might be a minor issue. Using machine learning and deep learning for skin cancer classification can increase accessibility and reduce the discomforting procedures involved in the traditional lesion detection process. These models can be wrapped in web or mobile apps and serve a greater population. In this paper, two such models are tested on the benchmark HAM10000 dataset of common skin lesions. They are Random Forest with Stratified K-Fold Validation, and MobileNetV2 (throughout the rest of the paper referred to as MobileNet). The MobileNet model was trained separately using both TensorFlow and Pytorch frameworks. A side-by-side comparison of both deep learning and machine learning models and a comparison of the same deep learning model on different frameworks for skin lesion diagnosis on a resource-constrained mobile environment has not been conducted before. The results indicate that each of these models fares better at different classification tasks. For greater recall, accuracy and detecting malignant melanoma, the Tensorflow Mobilenet was the better choice. However, for detecting noncancerous skin lesions, the Pytorch Mobilenet proved to be better. Random Forest was a better algorithm when it came to having a low computational cost with moderate accuracy.

# 1. Introduction to Research Background and Existing Research

Both in the U.S.A and around the world, Skin cancer has recently become one of the most prevalent types of cancer [1-5]. Skin diseases have been growing into a significant issue due to the inefficiency when it comes to diagnosing them in their early stages[6]. For instance, according to the American Skin Cancer Foundation, 2 people die from skin cancer every hour[14]. This can be preventable, as patients diagnosed early have a 95% chance of being cured[7].

Some common skin lesions include Squamous cell carcinoma, melanoma, intraepithelial carcinoma, and basal cell carcinoma. The most dangerous and cancerous of these is melanoma, with over 7230 deaths in the U.S.A in 2019[6]. Dermatoscopy is the current technique used in most clinics to diagnose skin cancer. Conducted by a dermoscope, this technique allows for the examination of skin lesions without any obstruction from skin surface reflections. Through dermoscopy, the accuracy of skin lesions treatment will be 75%-84% [9-11], but this process is human-labor intensive. Suspicious regions may even need to be magnified or illuminated to be seen more clearly [9, 12]. Some procedural algorithms that strive to improve dermoscopy include the ABCD-rule (Asymmetry, Border, Irregularity, Color variation, and Diameter), 3-point checklist, 7-point checklist, and the Menzies method[10, 13]. Regardless of these additions, dermoscopic imaging has considerable flaws due to the need for significant experience observing difficult situations, and the fine differences between skin lesions. Figure 1. illustrates the complexities of analyzing skin lesions by depicting melanoma and melanocytic nevi, two different conditions with strong visual resemblances. The relationship between these two lesions is further complicated by the fact that 33% of melanoma comes from developments in melanocytic nevi. Although melanocytic nevi form because of the BRAF V600E mutation of the BRAF gene, many will stay clinically stable and not progress to becoming cancerous [15].

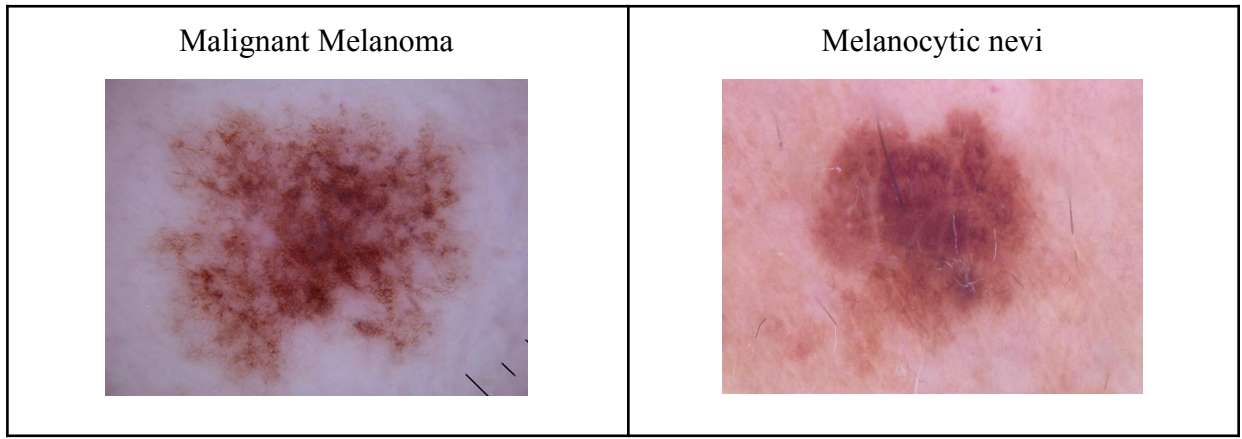

Figure 1: Side by side comparison of Malignant Melanoma and Melanocytic nevi (noncancerous growth)

In an attempt to solve the issues of dermoscopy being human-labor intensive, and hard to master even for the most experienced doctors[7], machine learning algorithms were developed to diagnose skin diseases based on images starting in the early 2000s[16].

Recently, deep learning models, starting with AlexNet in 2012[18], have been trained for classifying and segmenting skin lesions. Pre-trained CNNs, such as VGG, InceptionNet, DenseNet, and MobileNet have already been used on skin disease data and produced promising results. By pre-trained, this means that some degree of transfer learning was applied to these models, with preliminary training mostly on another dataset usually containing natural images. Here are some specific examples of previous research in this field:

In 2004, a probabilistic feature extraction model with a feedforward neural network was suggested by Sigurdsson et al. This method takes advantage of the molecular structure of skin lesions by using a nonlinear neural network and feature extraction based on Raman Spectroscopy[17].

She et al. attempted to combine skin line direction with the ABCD method (Asymmetry, Border irregularity, Color variegation, and Diameter of lesion) to classify benign and malignant lesions for Computer Automatic Diagnosis in 2007[16].

In 2015, Codella et. al. proposed a deep learning residual model that would be upgraded by class weight[19].

Dilated Convolution on pre-trained deep learning models, such as VGG16, VGG19, and InceptionV3 were implemented to improve accuracy while preserving the same computational complexity as traditional CNNs [7] in 2020.

In 2021, MIT's Institute for medical engineering and science attempted to use machine learning algorithms to detect suspicious pigmented lesions and implemented a novel approach to rank by lesion saliency[6].

Some Artificial Intelligence (AI) models have proven to have as good as or better accuracy than board-certified Dermatologists[20-24]. These include the Deep Convolutional Neural Network(DNN) proposed by Fujisawa et al[22] and Liu et. al's ResNet CNN which was used to classify melanomas and atypical nevi[24]. This high accuracy is not meant to replace dermatologists, but rather to expedite the diagnosis and detection of skin lesions. A notable number of dermatologists believe that AI can help with early diagnosis and decision-making[26]. The Deloitte 2018 Surveys of U.S. Healthcare Consumers and Physicians reported that anywhere between 58% to 69% of overall physicians, including dermatologists, wanted to increase their digital tool use[25]. This data point entails the use of AI. Furthermore, apps such as VisualDx, a skin lesion detection app specialized for pigmented skin, have been introduced to over 2300 Hospitals to assist diagnosis, with a 2019 survey asserting that dermatologists using the app saved about 22 minutes per day[26].

In this paper, we will use both machine learning and deep learning techniques to classify 7 common skin diseases and compare them based on their recall and how computationally efficient they are for a mobile environment. We will also compare the performance of the same deep learning model on different frameworks for the same task. Such a comparison has not been done before for a skin lesion dataset.

Here are the 7 diseases that will be classified. Figure 2 displays each of the diseases in more detail:

1. akiec - Actinic Keratoses (Solar Keratoses) and Intraepithelial Carcinoma (Bowen's disease). Actinic Keratoses (Solar Keratoses) and Intraepithelial Carcinoma are forms of precancers. They manifest through rough, scaly, and commonly unpigmented patches on the skin [8]. If not treated properly, they may deteriorate to become Squamous Cell Carcinoma - cancer that usually is unpigmented. Both diseases are commonly caused by extensive exposure to the sun, but Bowen's disease may be triggered by the Human Papillomavirus[27].

2. bcc - Basal Cell Carcinoma is the most common skin disease. If treated early, it inflicts minimal damage. Some signs of it include abnormal open sores, red patches, or growth on slightly elevated areas of the skin[6]. The characteristics of the lesion vary from person to person, with pigmentation being a significant distinction between darker-skinned and lighter-skinned people. Treatment progresses to becoming more elaborate depending on how severe the condition is but usually involves Mohs surgery [28].

3. bkl - benign keratosis-like lesions consist of 3 subclasses: seborrheic keratoses ("senile wart"), solar lentigo, and lichen-planus-like keratoses (LPLK). Although these subclasses may look different, they are all classified the same histopathologically [8]. These are noncancerous growths likely caused by age or extensive exposure to the sun and at times genetic factors. A challenge with lichen planus-like keratosis is that it may exhibit a structure similar to that of melanoma, and thus is often biopsied[29].

4. df- Dermatofibroma. Noncancerous skin nodules. Occurs more commonly in women and the lower extremities. The most common pattern involves a white patch in a center with a pigment network[30].

5. nv- Melanocytic nevi. Noncancerous skin lesions. Usually more symmetrical in terms of color and structure compared to Melanoma[31]. Exposure to the sun is thought to cause onset. In layman's terms, it is known as the birthmark or mole [8]. People with multiple nevi are at more risk for developing cancerous moles known as melanoma[32].

6. mel- Melanoma. Is the malignant form of a mole. May be invasive-spreading to other areas of the body other than the starting tissue- or noninvasive. If detected early, treatment includes removal by surgical excision. Symptoms include a change in an existing mole[33].

7. vasc - Vascular Skin Lesions are commonly known as birthmarks. They include cherry angiomas, angiokeratomas and pyogenic granulomas. They are common benign skin tumors. Cherry angiomas are common skin growths that are red due to the collection of blood vessels around them[34]. Pyogenic granulomas are classified as red bumps commonly forming after small injuries; these growths are prevalent in women and children[35]. Hemorrhage is also included in this category, which is bleeding that accumulates on the skin due to broken blood vessels.

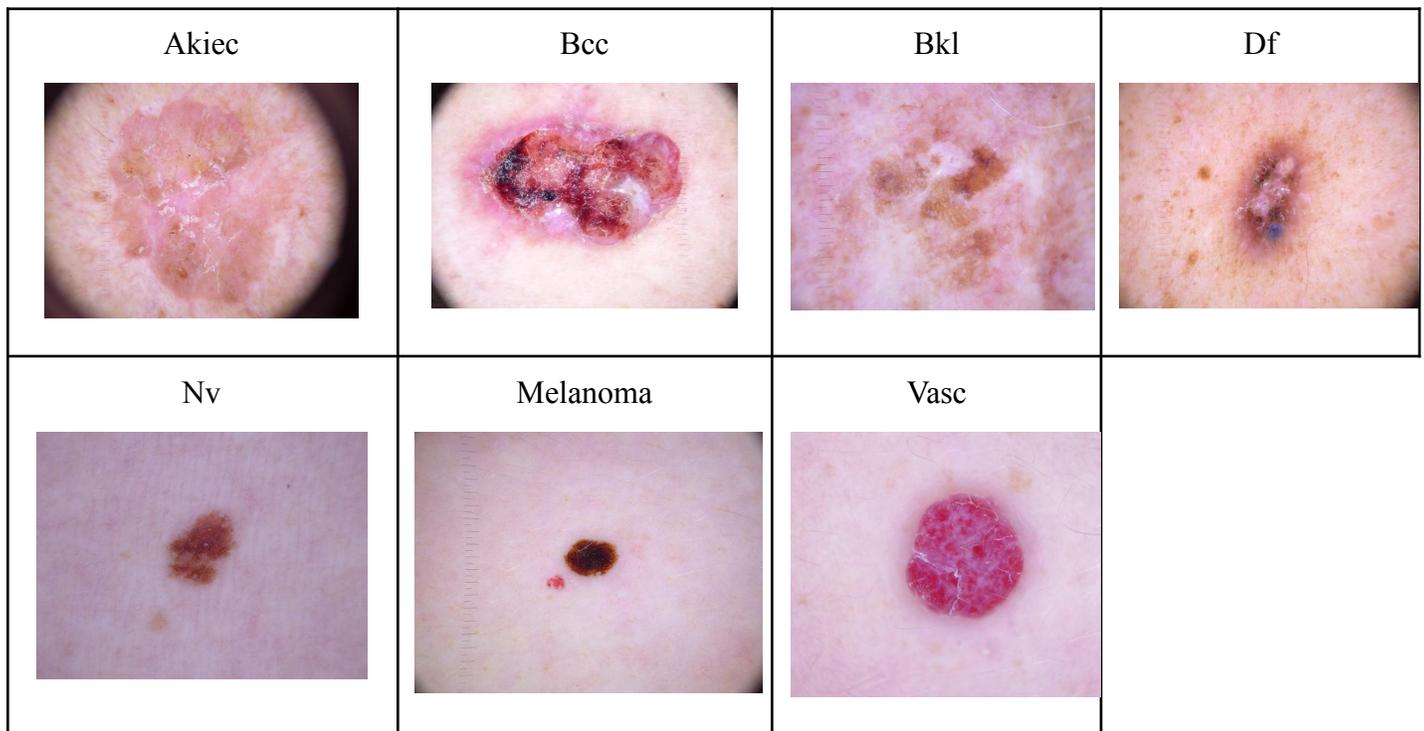

Figure 2: 7 Images representing each of the classes from the HAM10000 dataset: Actinic keratoses, Basal cell carcinoma, Benign keratosis-like lesions, Dermatofibroma, Melanocytic Nevi, Vascular lesions, Melanoma

Melanoma and Basal Cell Carcinoma are cancerous, Actinic Keratoses has the potential to be cancerous, and the other skin lesions are benign.

**2. Research Materials: Dataset and Exploratory Data Analysis**

The HAM10000 dataset, also referred to as the International Skin Imaging Collaboration (ISIC) 2018 dataset, includes 10015 images with 7 classes (see Figure 3). There is a drastic data imbalance, with much more Melanocytic nevi images than any other class. These 7, generic classes were included because a model trained with this data was only supposed to serve as a benchmark for skin lesion classification[8].

The dataset was collected from two different sites, the Department of Dermatology at the Medical University of Vienna, Austria, and the skin cancer practice of Cliff Rosendahl in Queensland, Australia. They were collected over a span of 20 years. PowerPoint files and Excel databases were used by the Australian location to store images and meta-data. The Medical

University of Vienna utilized multiple formats to store images and metadata, since they started gathering these images before digital cameras became common [8].

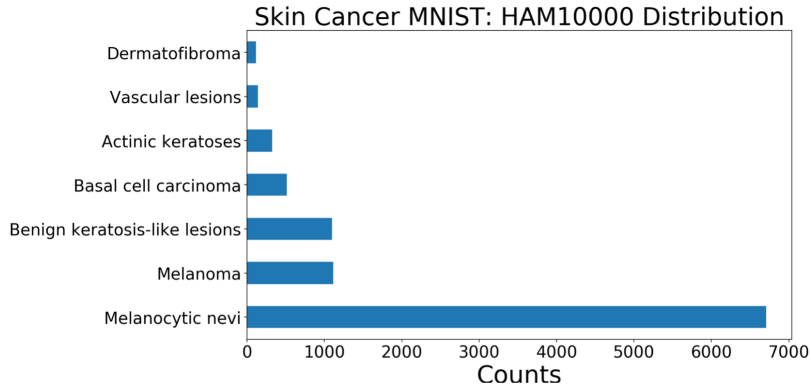

Figure 3: Distribution of Images in the Dataset

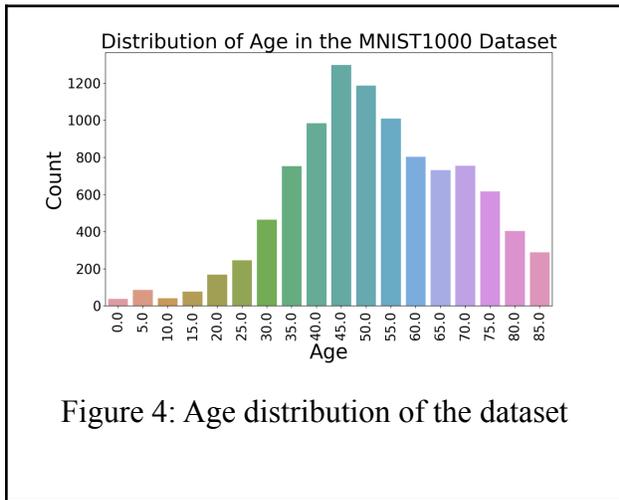

Figure 4: Age distribution of the dataset

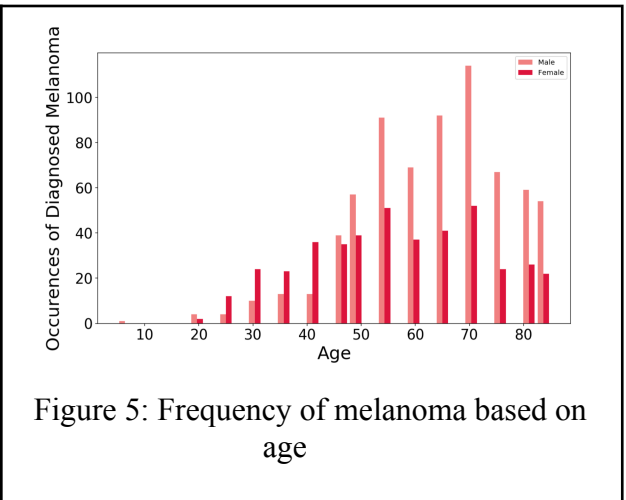

Figure 5: Frequency of melanoma based on age

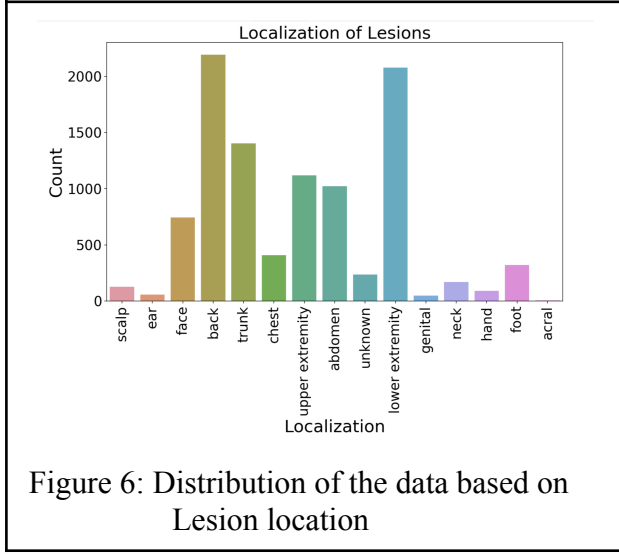

Figure 6: Distribution of the data based on Lesion location

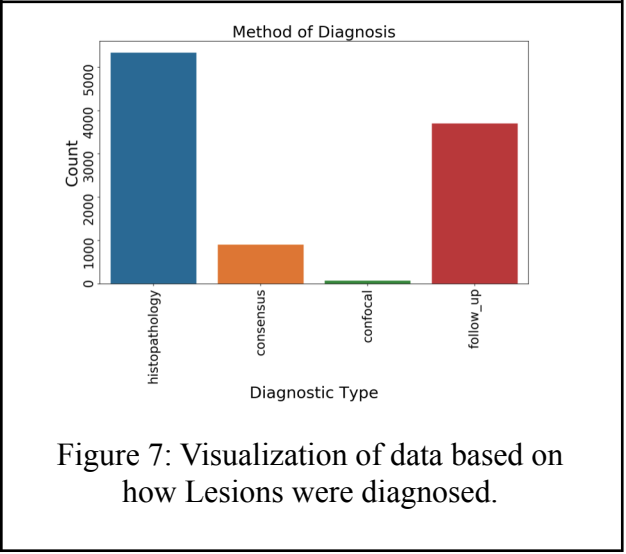

Figure 7: Visualization of data based on how Lesions were diagnosed.

In terms of other distributions based on categories, only 16% of the total images are cancerous, and the rest are non-cancerous. Figures 4 and 5 show the data is largely collected from people in their middle ages (40-50), with more data from males than women. Figure 5 supports this by showing that there is more occurrence of melanoma in men than women in the population after 50. This is because men are perhaps more prone to developing melanoma after age 49 compared to women, which is where this dataset is concentrated[41]. According to the American Academy of Dermatology, men, in general, are less aware of factors that lead to cancer. These include facts like a base tan does not protect from being harmed by the sun's rays. From an anatomic perspective, a man's skin contains more collagen and estasin and is also thicker and has less fat below it than a woman's. This structure makes a man's skin more susceptible to being harmed by UV light.

A large number of the lesion data comes from the lower extremity and back of the body as seen in Figure 6. Figure 7 represents how most of the datasets were technically validated. The methods for this included histopathology, which is examining tissue from the lesion through a biopsy; confocal images which is a microscopic imaging technique, follow-up which was unique to nevi and declared the lesion to be benign if did not show any changes during 3 follow-up visits or 1.5 years; and Consensus which involved expert consensus rating for benign entries. The majority of the data has been verified by histopathology as a ground truth (Figure 7).

## 3. Algorithms:
### 3.1 Random forest (machine learning algorithm) :

Random forest was used on the dataset as a preliminary baseline model. Since the HAM10000 is a multiclass, categorical, and unbalanced dataset, random forest was thought to perform better than other machine learning models such as logistic regression and SVM [43]. Random forest is a machine learning algorithm that employs an ensemble of uncorrelated decision trees to predict an output[37]. Each tree will produce its class prediction, and the total predictions will be tallied as votes, with the most frequently predicted class as the final output [36].

KFold Cross Validation with Stratification [42] is applied to assess how the dataset can perform on unseen data. It involves breaking the dataset into k-folds, with one fold representing the validation while the rest representing the training data. This technique will provide a less biased outcome compared to just splitting the dataset into train and test. In addition, stratified K-fold is used to keep the same proportion of classes in each fold as that of the original dataset, something that is essential for an unbalanced dataset like HAM10000 .

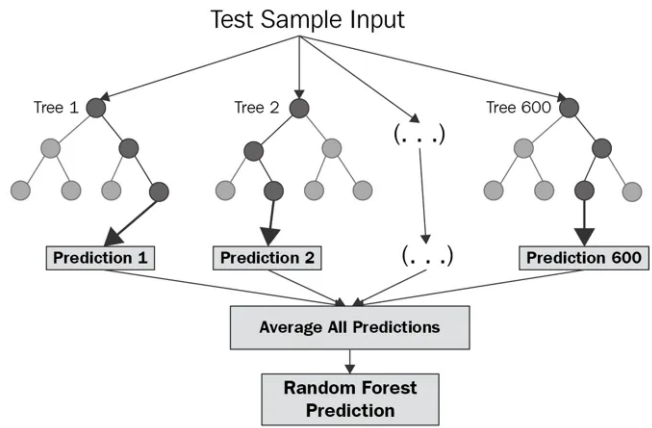

Figure 8: Illustration of Random Forest Architecture[44]

**3.2 Mobilenet algorithm (deep learning algorithm models):**

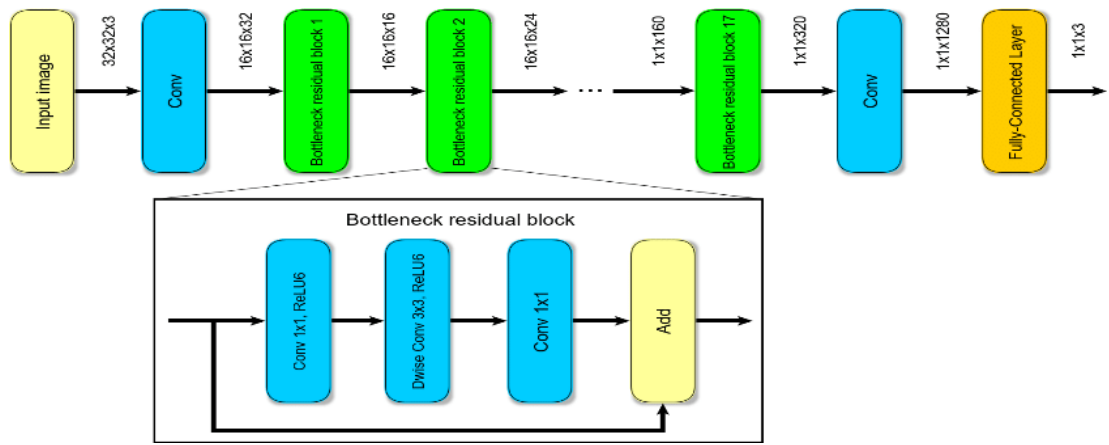

Figure 9: Architecture of MobileNetV2[45]

MobilenetV2 was tried on the dataset because it is the most recent model presented by Google. Mobilenet serves as a computer vision model that could operate with high accuracy in the resource-constrained mobile environment[38]. It is deployable to mobile devices, which is important for the issue of skin classification, as it allows for greater accessibility and portability for diagnosis. It achieved this by using depthwise-separable convolution. MobileNetV2 improved on the original MobileNet by adding Linear Bottlenecks between the layers and shortcut connections between these bottlenecks. The reasoning behind this was that the shortcuts allow for better training and accuracy, and the bottleneck layers would increase computation per module, but decrease the number of layers[39]. The key attributes of this model are that it is an inverted residual model, with residuals between its bottleneck layers, that strives to be more compact, yet accurate compared to other models. Another aspect of the model is its use of Depthwise Separable Convolution[39]. Since this entails splitting the 3d convolution of an image into two parts, a depthwise convolution and a pointwise convolution, it can achieve the convolution with more computational efficiency than regular convolution.

The Tensorflow and Pytorch versions of the same model were used to evaluate which one would be best suited for the skin classification task. The main difference between these two models is the way they execute and process code. In Pytorch, one can seamlessly access the GPUs that run on CUDA to further enhance the training of a dataset [47]. On the other hand, the GPU cannot be as easily accessed on TensorFlow since it uses its own inbuilt data [46]. PyTorch is also similar to the python programming language providing a native approach, whereas TensorFlow at times seems completely different from python.

**4. Description of Methodology:**
**4.1 Data Pre-Processing Augmentation:**

The dataset was augmented for both TensorFlow and PyTorch MobileNets. For both, the images were resized to 224 by 224 images and normalized. Rotations of 180 degrees, translations in the x and y direction of the figure size and random horizontal and vertical flips were used to augment the data.

**4.2 Training Approach:**

Random forest is considered a baseline model for classification, it was trained on the CPU. The number of trees used was 40. Stratified K-Fold was applied with 10 folds.

MobilenetV2 was trained separately on both Tensorflow and PyTorch. For the TensorFlow model, the last 5 layers were deleted and a new dense layer with the output changed to 7 elements instead of the original 1000 elements to account for the 7 output classes. Both were pre-trained on the ImageNet dataset. Both were also trained on the NVIDIA Tesla K80 GPU for 1000 epochs.

The learning Rate for the Pytorch Mobilenet was 0.001. The rate decreased by a factor of 0.1 every 7 epochs due to an exp_lr_scheduler.

## 5. Results and conclusions:

In terms of accuracy, the MobileNet TensorFlow Model had the best results followed by MobileNet on PyTorch and finally Random forest. The Tensorflow MobileNet yielded the greatest accuracy with 86.5% followed by the Pytorch MobileNet at 78.3% and the Random Forest at 74.9%. Although the TensorFlow MobileNet model did provide higher accuracy than the other two models, this metric is largely irrelevant due to the dataset being imbalanced. A better understanding of which model is better can be gained from a class-by-class analysis. We prioritize the recall metric while conducting this comparison because in a medical dataset like HAM10000, it needs to keep track of how many of the true positives were found rather than the accuracy of positive predictions (precision).

Each of the 3 models classified melanocytic nevi with the highest recall: 100%, 95%, and 93% for Random Forest, TensorFlow Mobilenet, and Pytorch MobileNet respectively. This proves that in general, all 3 models can classify nevi with great accuracy. In this case, either the Tensorflow model may be the one to choose based on its higher accuracy, or the Random Forest due to its computational simplicity and relatively substantial results.

As for the worst-performing class, for the Pytorch MobileNet with 33% and the Random Forest Algorithm with 0.00%, it was dermatofibroma. This can be attributed to the fact that there were drastically fewer dermatofibroma images compared to other classes. The lowest performing class for MobileNet with Tensorflow was benign keratosis-like lesions with a 41% recall.

For melanoma, the deadliest class in the dataset, the Tensorflow MobileNet yielded better results with a recall of 64% compared to 14% for the Random Forest and 45% for the PyTorch MobileNet.

On the other hand, for non-cancerous lesions, which are more likely to be effectively treated and given adequate care through a cellphone app, the Pytorch Tensorflow showed a better outcome; it had a recall of 52% on actinic keratoses and 86% on vascular skin lesions. As a comparison, the Pytorch MobileNet had 42% for Actinic Keratoses and 55% for Vascular Skin Lesions, and Random Forest had 18% for Actinic Keratoses and 0% for the latter.

**5.1 Accuracy Precision Recall F-1 Score:**

Since the HAM10000 dataset is heavily imbalanced, evaluation measures such as precision, recall, and F-1 score must be relied on to better evaluate the models.

Figures 10, 11, and 12 display the precision, recall, and F-1 score for the Random Forest, MobileNet with TensorFlow, and Mobilenet with Pytorch respectively.

|  | **Precision** | **Recall** | **F-1 Score** | **Support** |
|---|---|---|---|---|
| **akiec** | 0.60 | 0.18 | 0.28 | 33 |
| **bcc** | 0.30 | 0.18 | 0.22 | 51 |
| **bkl** | 0.55 | 0.26 | 0.36 | 110 |
| **df** | 0.00 | 0.00 | 0.00 | 11 |
| **mel** | 0.53 | 0.14 | 0.23 | 111 |
| **nv** | 0.76 | 1.00 | 0.86 | 671 |
| **vasc** | 0.00 | 0.00 | 0.00 | 15 |
| **Accuracy** |  |  | 0.73 | 1002 |
| **Macro Avg** | 0.39 | 0.25 | 0.28 | 1002 |
| **Weighted Avg** | 0.66 | 0.73 | 0.66 | 1002 |

Figure 10: Precision, Recall, F-1-score for validation set for Random Forest

|  | Precision | Recall | F-1 Score | Support |
|---|---|---|---|---|
| **akiec** | 0.65 | 0.42 | 0.51 | 26 |
| **bcc** | 0.69 | 0.80 | 0.74 | 30 |
| **bkl** | 0.63 | 0.41 | 0.50 | 75 |
| **df** | 0.50 | 0.50 | 0.50 | 6 |
| **mel** | 0.35 | 0.64 | 0.45 | 39 |
| **nv** | 0.94 | 0.95 | 0.95 | 751 |
| **vasc** | 0.86 | 0.55 | 0.67 | 11 |
| **Accuracy** |  |  | 0.86 | 938 |
| **Macro Avg** | 0.66 | 0.61 | 0.62 | 938 |
| **Weighted Avg** | 0.87 | 0.86 | 0.86 | 938 |

Figure 11: Precision, Recall, F-1-score for test set of TensorFlow MobileNet

|  | Precision | Recall | F-1 Score | Support |
|---|---|---|---|---|
| **akiec** | 0.31 | 0.52 | 0.39 | 33 |
| **bcc** | 0.50 | 0.61 | 0.55 | 51 |
| **bkl** | 0.69 | 0.45 | 0.55 | 110 |
| **df** | 0.18 | 0.33 | 0.24 | 12 |
| **mel** | 0.57 | 0.45 | 0.50 | 111 |
| **nv** | 0.92 | 0.93 | 0.92 | 670 |
| **vasc** | 0.38 | 0.86 | 0.52 | 14 |
| **Accuracy** |  |  | 0.78 | 1001 |
| **Macro Avg** | 0.51 | 0.59 | 0.52 | 1001 |
| **Weighted Avg** | 0.80 | 0.78 | 0.79 | 1001 |

Figure 12: Precision, Recall, F-1-score for test set of PyTorch MobileNet

## 5.2 Confusion Matrix

Confusion matrices are used to further analyze how well each model is doing. This provided more insight into where the model was making the most mistakes and confusing classes. For instance, in the Random Forest Classifier, the model tended to mix up actinic keratoses with nevi, and the PyTorch model tended to mistake dermatofibroma as basal cell carcinoma. Figures 13, 14 and 15 display these intricacies in more detail.

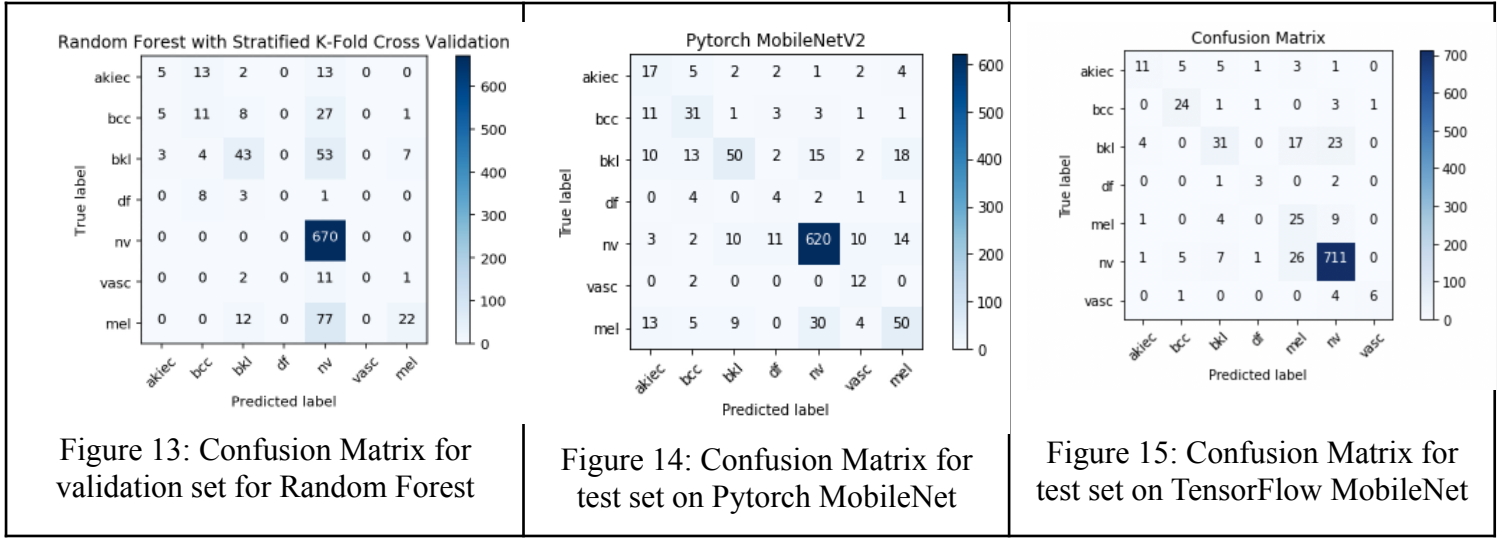

Figure 13: Confusion Matrix for validation set for Random Forest

Figure 14: Confusion Matrix for test set on Pytorch MobileNet

Figure 15: Confusion Matrix for test set on TensorFlow MobileNet

## 6. Discussion

The final results were analyzed based on average F-1 score, recall, and precision. The recall was given the most importance. Precision represents the number of positive classifications that were correct whereas recall is a representation of what percentage of the actual positive class was correctly identified. Hence, recall is prioritized as there is more grace for classifying something like cancerous melanoma as cancer even when it is not compared to not identifying an actual melanoma case. Once again, this is supposed to be an early diagnosis model.

A key concern for incorporating this model in a mobile app is deployability. Tensorflow is more well-established for web and mobile platforms, providing options such as Tensorflow Lite. PyTorch just started to provide more mobile deployment options such as PyTorch lite[46]. PyTorch Mobile is more convenient because PyTorch Mobile and PyTorch are the same framework and share the same codebase, whereas TFLite and TensorFlow are different frameworks that share the same name. Some TensorFlow models cannot be converted to TFLite models.

In terms of classification, each model had distinct shortcomings as well. For example:
- Random Forest was unable to classify none of the dermatofibroma or vascular skin lesions correctly due to the meager amounts of data for these two diseases in the dataset (both have less than 200 images)
- The TensorFlow MobileNet model performed poorly on a multitude of noncancerous skin lesions such as actinic keratoses with a 42% recall, and vascular skin Lesions at 55%. It fared better than the PyTorch model for cancerous lesions such as melanoma by a large margin, a 19% difference in the recall.
- On the contrary, MobileNet on PyTorch produced promising results on non-cancerous skin lesions, 52% on Actinic Keratoses and 86% on Vascular Skin lesions, but did not do as well on cancerous growths.
- Although the TensorFlow MobileNet model fared better overall, PyTorch does have notable advantages such as more easy access to the GPU and a more native approach for python developers

Each model has distinct advantages that make it better than the others. When looking for a model that diagnoses cancerous skin lesions with great accuracy, MobileNetV2 with TensorFlow may be the best option, whereas if non-cancerous lesion accuracy is prioritized, MobileNet with PyTorch serves better. If computational cost is a concern, Random Forest can be more suitable as it still provides a relatively competitive accuracy at a lower computational cost. In terms of the ultimate goal which would be deploying this model on an easily accessible mobile application, the MobileNet CNNs should be given preference, since the whole purpose of the model was to function in highly constrained mobile environments while ensuring high accuracy.

Hopefully, this novel analysis from a mobile of deep learning and machine learning algorithms and frameworks based on computational capability for mobile environments will provide helpful insight for medical advancements dealing with the pressing problem of efficient early diagnosis of skin lesions. Further work includes fine-tuning and deploying the model onto a mobile application to make early diagnosis more accessible.

**Bibliography:**
[1] Codella, N. C., Gutman, D., Celebi, M. E., Helba, B., Marchetti, M. A., Dusza, S. W., ... & Halpern, A. (2018, April). Skin lesion analysis toward melanoma detection: A challenge at the 2017 international symposium on biomedical imaging (isbi), hosted by the international skin imaging collaboration (isic). In 2018 IEEE 15th International Symposium on Biomedical Imaging (ISBI 2018) (pp. 168-172). IEEE.
[2] R. M.- Cancer and undefined 1995, "An overview of skin cancers," *Wiley Online Libr*.
[3] Rogers, H. W., Weinstock, M. A., Feldman, S. R., & Coldiron, B. M. (2015). Incidence estimate of nonmelanoma skin cancer (keratinocyte carcinomas) in the US population, 2012. JAMA dermatology, 151(10), 1081-1086.
[4] C. Facts, "Cancer Facts & Figures 2017 Cancer Facts & Figures 2017 Special Section : Rare Cancers in Adults (/ content / dam / cancer- Cancer Facts & Figures 2017," pp. 1–7, 2017.
[5] Siegel, R. L., Miller, K. D., & Jemal, A. (2015). Cancer statistics, 2015. CA: a cancer journal for clinicians, 65(1), 5-29.
[6] Soenksen, Luis R., et al. "Using deep learning for dermatologist-level detection of suspicious pigmented skin lesions from wide-field images." Science Translational Medicine 13.581 (2021).
[7] Ratul, Md Aminur Rab, et al. "Skin lesions classification using deep learning based on dilated convolution." BioRxiv (2020): 860700.
[8]Tschandl, P., Rosendahl, C. & Kittler, H. The HAM10000 dataset, a large collection of multi-source dermatoscopic images of common pigmented skin lesions. *Sci Data* 5, 180161 (2018). https://doi.org/10.1038/sdata.2018.161
[9] Kittler, H., Pehamberger, H., Wolff, K., & Binder, M. (2002). Diagnostic accuracy of dermoscopy. The lancet oncology, 3(3), 159-165.
[10] Binder, M., Schwarz, M., Winkler, A., Steiner, A., Kaider, A., Wolff, K., & Pehamberger, H. (1995). Epiluminescence microscopy: a useful tool for the diagnosis of pigmented skin lesions for formally trained dermatologists. Archives of dermatology, 131(3), 286-291.
[11] Carli, P & Quercioli, E & Sestini, Serena & Stante, M & Ricci, L & Brunasso, G & De Giorgi, Vincenzo. (2003). Pattern analysis, not simplified algorithms, is the most reliable method for teaching dermoscopy for melanoma diagnosis to residents in dermatology. The British journal of dermatology. 148. 981-4. 10.1046/j.1365-2133.2003.05023.x.
[12] Argenziano, G., Soyer, H. P., Chimenti, S., Talamini, R., Corona, R., Sera, F., ... & Hofmann-Wellenhof, R. (2003). Dermoscopy of pigmented skin lesions: results of a consensus meeting via the Internet. Journal of the American Academy of Dermatology, 48(5), 679-693.
[13] Gachon, J., Beaulieu, P., Sei, J. F., Gouvernet, J., Claudel, J. P., Lemaitre, M., ... & Grob, J. J. (2005). First prospective study of the recognition process of melanoma in dermatological practice. Archives of dermatology, 141(4), 434-438
[14] "Skin Cancer Facts & Statistics." The Skin Cancer Foundation, 10 Aug. 2021, www.skincancer.org/skin-cancer-information/skin-cancer-facts/#:~:text=More%20than%202%20people%20die,for%20melanoma%20is%2099%20percent.
[15] Damsky, W. E., and Marcus Bosenberg. "Melanocytic nevi and melanoma: unraveling a complex relationship." Oncogene 36.42 (2017): 5771-5792.
[16] She, Z., Liu, Y., & Damatoa, A. (2007). Combination of features from skin pattern and ABCD analysis for lesion classification. Skin Research and Technology, 13(1), 25-33.
[17] Sigurdsson, S., Philipsen, P. A., Hansen, L. K., Larsen, J., Gniadecka, M., & Wulf, H. C. (2004). Detection of skin cancer by classification of Raman spectra. IEEE transactions on biomedical engineering, 51(10), 1784-1793.

[18] Esteva, A., Kuprel, B., & Thrun, S. (2015). Deep networks for early stage skin disease and skin cancer classification. Project Report, Stanford University

[19] Codella, N., Cai, J., Abedini, M., Garnavi, R., Halpern, A., & Smith, J. R. (2015, October). Deep learning, sparse coding, and SVM for melanoma recognition in dermoscopy images. In International workshop on machine learning in medical imaging (pp. 118-126). Springer, Cham.

[20] T. J. Brinker, A. Hekler, A. H. Enk, J. Klode, A. Hauschild, C. Berking, B. Schilling, S. Haferkamp, D. Schadendorf, S. Fröhling, J. S. Utikal, C. Kalle, A convolutional neural network trained with dermoscopic images performed on par with 145 dermatologists in a clinical melanoma image classification task. Eur. J. Cancer 111, 148–154 (2019).

[21] J. Zhang, Y. Xie, Y. Xia, C. Shen, Attention residual learning forskin lesion classification. IEEE Trans. Med. Imaging 38, 2092–2103 (2019).

[22] Y. Fujisawa, Y. Otomo, Y. Ogata, Y. Nakamura, R. Fujita, Y. Ishitsuka, R. Watanabe, N. Okiyama, K. Ohara, M. Fujimoto, Deep-learning-based, computer-aided classifier developed with a small dataset of clinical images surpasses board-certified dermatologists in skin tumour diagnosis. Br. J. Dermatol. 180, 373–381 (2019).

[23] X.-Y. Zhao, X. Wu, F.-F. Li, Y. Li, W.-H. Huang, K. Huang, X.-Y. He, W. Fan, Z. Wu, M.-L. Chen, J. Li, Z.-L. Luo, J. Su, B. Xie, S. Zhao, The application of deep learning in the risk grading of skin tumors for patients using clinical images. J. Med. Syst. 43, 283 (2019).

[24] Y. Liu, A. Jain, C. Eng, D. H. Way, K. Lee, P. Bui, K. Kanada, G. deOliveira Marinho, J. Gallegos, S. Gabriele, A deep learning system for differential diagnosis of skin diseases. Nat. Med., 1–9 (2020).

[25] Palmer, Whitney J. "Digital Tools Impact DAILY Dermatology Practice." *Dermatology Times*, Dermatology Times, 13 Nov. 2020, www.dermatologytimes.com/view/digital-tools-impact-daily-dermatology-practice.

[26] Art Papier, M.D., and FAAD Steve Xu M.D. "Ai: It's Not Dermatologist vs. Machine." *Dermatology Times*, Dermatology Times, 13 Nov. 2020, www.dermatologytimes.com/view/ai-its-not-dermatologist-vs-machine.

[27] "Actinic Keratosis." *Mayo Clinic*, Mayo Foundation for Medical Education and Research, 13 Jan. 2021, www.mayoclinic.org/diseases-conditions/actinic-keratosis/symptoms-causes/syc-20354969.

[28] Ratner, Désirée. "Basal Cell Carcinoma." *The Skin Cancer Foundation*, 9 Aug. 2021, www.skincancer.org/skin-cancer-information/basal-cell-carcinoma/.

[29] Moscarella, E. et al. Lichenoid keratosis-like melanomas. *J Am Acad Dermatol* 65, e85, Van de (2011).

[30] Zaballos, P., Puig, S., Llambrich, A. & Malvehy, J. Dermoscopy of dermatofibromas: a prospective morphological study of 412 cases. *Arch Dermatol* 144, 75–83 (2008)

[31] Rosendahl, C., Cameron, A., McColl, I. & Wilkinson, D. Dermatoscopy in routine practice - 'chaos and clues'. *Aust Fam Physician* 41, 482–487 (2012).

[32] "Melanocytic Naevus." *Melanocytic Naevus | DermNet NZ*, dermnetnz.org/topics/melanocytic-naevus/.

[33] "What Is Melanoma Skin CANCER?: What Is Melanoma?" *American Cancer Society*, www.cancer.org/cancer/melanoma-skin-cancer/about/what-is-melanoma.html.

[34] Zaballos, P. et al. Dermoscopy of solitary angiokeratomas: a morphological study. *Arch Dermatol* 143, 318–325 (2007).

[35] Zaballos, P. et al. Dermoscopy of pyogenic granuloma: a morphological study. *Br J Dermatol* 163, 1229–1237 (2010).

[36] Brownlee, Jason. "A Gentle Introduction To k-Fold Cross-Validation." *Machine Learning Mastery*, 2 Aug. 2020, machinelearningmastery.com/k-fold-cross-validation/.

[37] Breiman, Leo. "Random forests." Machine learning 45.1 (2001): 5-32.

[38] Howard, Andrew G., et al. "Mobilenets: Efficient convolutional neural networks for mobile vision applications." *arXiv preprint arXiv:1704.04861* (2017).

[39] Sandler, Mark, et al. "Mobilenetv2: Inverted residuals and linear bottlenecks." *Proceedings of the IEEE conference on computer vision and pattern recognition*. 2018.

[40] Biau, Gérard, and Erwan Scornet. "A random forest guided tour." *Test* 25.2 (2016): 197-227. Matthews NH, Li WQ, Qureshi AA, et al. Epidemiology of Melanoma. In: Ward WH, Farma JM, editors. Cutaneous Melanoma: Etiology and Therapy [Internet]. Brisbane (AU): Codon Publications; 2017 Dec 21. Chapter 1. Available from: https://www.ncbi.nlm.nih.gov/books/NBK481862/ doi: 10.15586/codon.cutaneousmelanoma.2017.ch1

[41] *"Melanoma Strikes Men Harder." American Academy of Dermatology,* www.aad.org/public/diseases/skin-cancer/types/common/melanoma/men-50.

[42] Brownlee, Jason. "How to Fix k-Fold Cross-Validation for Imbalanced Classification." *Machine Learning Mastery*, 30 July 2020, machinelearningmastery.com/cross-validation-for-imbalanced-classification/.

[43] Muchlinski, D., Siroky, D., He, J., & Kocher, M. (2016). Comparing Random Forest with Logistic Regression for Predicting Class-Imbalanced Civil War Onset Data. *Political Analysis, 24*(1), 87-103. doi:10.1093/pan/mpv024

[44] Chakure, Afroz. "Random Forest Structure." *Medium.com*, 29 June 2019, miro.medium.com/max/700/0*f_qQPFpdofWGLQqc.png.

[45] Seidaliyeva, Ulzhalgas. "The Architecture of the MobileNetv2 Network." *Researchgate.net*, July 2020, www.researchgate.net/figure/The-architecture-of-the-MobileNetv2-network_fig3_342856036.

[46] Kurama, Vihar. "PyTorch vs. TensorFlow: Which Framework Is Best for Your Deep Learning Project?" *Built In*, 9 Sept. 2020, builtin.com/data-science/pytorch-vs-tensorflow.

[47] Dubovikov, Kirill. "PyTorch vs TensorFlow - Spotting the Difference." *Medium*, Towards Data Science, 15 Oct. 2018, towardsdatascience.com/pytorch-vs-tensorflow-spotting-the-difference-25c75777377b.